# Addressing Behavioral Change towards Energy Efficiency in European Educational Buildings

Georgios Mylonas
D. Amaxilatis
Comp. Tech. Inst. &
Press "Diophantus",
Patras, Greece
mylonasg@cti.gr

H. Leligou, T.Zahariadis,
E. Zacharioudakis
Synelixis Solutions Ltd.
Chalkida, Greece
nleligou@synelixis.com

J. Hofstaetter,
A. Friedl
OVOS
Vienna, Austria
{jho, af}@ovos.at

F. Paganelli,
G. Cuffaro
CNIT
Florence, Italy
{federica.paganelli,
giovanni.cuffaro}@cnit.it

Jimm Lerch

Eurodocs
Soderhamn, Sweden
jimm@eurodocs.net

*Abstract*—Energy consumption reserves a large portion of the budget for school buildings. At the same time, the students that use such facilities are the adults of the years to come and thus, should they embrace energy-aware behaviors, then sustainable results with respect to energy efficiency are anticipated. GAIA is a research project targeting this user domain, proposing a set of applications that a) aims at raising awareness, prompting action and fostering engagement in energy efficiency enhancement, and b) is adaptable to the needs of each facility/community. This application set relies on an IoT sensing infrastructure, as well as on the use of humans as sensors to create situational awareness.

I. INTRODUCTION

Of the primary energy consumption in the EU, buildings are responsible for about 40%, while most of the energy is directed towards heating and cooling systems. By improving the energy efficiency of residential and public buildings, the EU's total energy consumption can be reduced by about 6% and lower $CO_2$ emissions by about 5%, [1], [2]. However, reducing energy use in buildings by introducing new technologies is very challenging; across Europe, the rates of construction of new buildings as well as the rates of renovation of existing ones are generally very low [3].

In order to achieve the energy reduction targets set for a sustainable future, affecting the behavioral characteristics of the citizens' interaction within the buildings where they live and work, will have a more tangible impact [4]. Furthermore, focusing on a particular target group is essential for designing a successful behavioral-change strategy [5]. Educational buildings constitute 17% of the non-residential building stock in the EU [3], with energy expenses in schools have been treated as relatively fixed and inevitable. Evidence shows that a focus on energy use in schools yields an array of important rewards in concert with educational excellence and a healthy learning environment [6]. GAIA[1] is an ongoing EU research project focused on the educational community, aiming to instill behavioral change towards energy efficiency and sustainability through an educational approach, combined with a set of applications, described in this work, supporting such aspects.

Targeting energy efficiency in the context of the educational community can yield significant results because: (a) educating children and young people to embrace energy-efficient habits has sustainable results on energy consumption, as these people are highly unlike to abolish their habits in the future, (b) focusing on students will also affect their immediate family environment. Several studies document the ability of students to influence choices made by their families related to environmental issues [8]. Research interviews conducted [9] made clear that energy conservation insights learned in school can be applied at home by students and their families. In the EU, people aged under 30 represent about a third of the total population [7], thus a large part of the population can be affected. This approach is further strengthened by actively involving school staff (e.g., school directors, building managers), which can offer critical insights about ways to lower a building's energy footprint [8], [10], [11].

To shape the behaviour of the school/universities' communities, GAIA develops tools for raising awareness, for motivating and prompting energy efficient behaviours and for fostering engagement to ensure a sustainable result. In the design of the toolset, a set of great challenges was faced: a) these communities include people of diverse ages which implies that different communication channels have to be supported. b) customization to the automation level of each school is mandatory c) localization of the application is crucial for the penetration of the applications and this does not refer only to language but also to the roles and availability of staff to implement energy efficiency campaigns; for example, while in Sweden and in independent schools in Greece the role of building manager is played by dedicated personnel, in public schools in Greece and in Italy, this role is undertaken by the school principals.

The rest of the paper is organized as follows: in section 2, we provide a short introduction to the architecture upon which GAIA's application set, described in this work, is based. In section 3, we present our approach to behaviour change and the high level description of the GAIA application set that implement our approach. Next, in section 4 we elaborate on the GAIA application set functionality and technical challenges, while in section 5 we provide a discussion on our approach, and then conclude this work in section 6.

---

[1] Green Awareness In Action website, http://gaia-project.eu/

## II. ARCHITECTURE OVERVIEW

The overall architecture of GAIA is built upon multiple IoT installations in school buildings. Each installation consists of a multitude of IoT nodes, which communicate with a cloud infrastructure via a gateway device. The IoT nodes include multiple sensing devices, while the gateway nodes coordinate communication and enable interaction with cloud-based services and other Internet-connected devices. All devices in our system use a wireless interface to communicate locally, while gateways use wired connectivity to report measurements and receive commands or configuration updates.

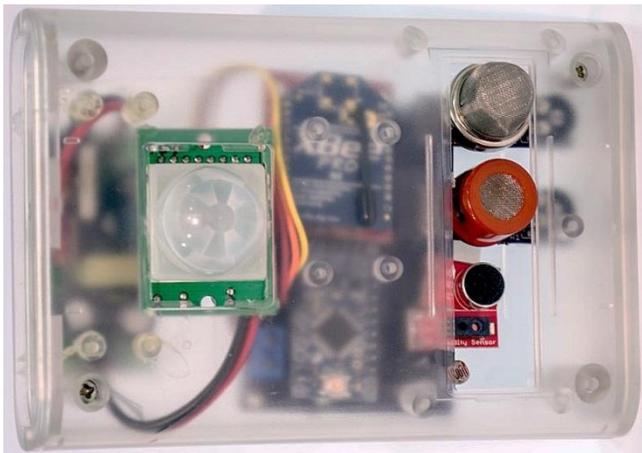

*Figure 1 The IoT node device used inside classrooms*

The overall design pattern for the installation of IoT infrastructure in the school buildings participating in the project, is as follows:

- IoT nodes installed to monitor the power consumption of the building as a whole, or specific floors/sectors (depending on the peculiarities of each building).
- IoT nodes installed in classrooms and other supporting rooms to monitor a set of environmental parameters such as temperature, humidity, activity and noise levels (see Fig. 1).
- A set of IoT gateway nodes installed in central points of the building to bridge the IoT nodes that communicate using IEEE 802.15.4 with the Internet. The IoT gateways communicate directly with GAIA cloud services.

The hardware design of the IoT nodes utilized in GAIA in almost their entirety follows an open-source approach, using hardware components widely available, paired with some custom designs, for example custom PCBs for interconnecting sensors to MCUs, etc. A large part of the available infrastructure are based on Arduino-compatible or Raspberry Pi components, with both being the platform of choice for semi-custom IoT hardware design for a large part of the research community. GAIA will release the specifications for this infrastructure and related results as open-source, in all cases wherever possible.

Based on this IoT infrastructure, GAIA continuously monitors energy consumption and environmental parameters inside school buildings that participate in the project. Such data will be used as input to the educational aspects of the project, as well as to the application set described in the next sections. The idea is to provide students and educators with a more "personalized" approach: GAIA's educational material and applications will use the actual situation in the school buildings to make interaction more engaging, fun and useful. The next sections provide more details on the overall design and concepts utilized in GAIA, aiming to produce such a result.

## III. BEHAVIORAL CHANGE APPROACH

To change the behavior and achieve sustainable results, we propose a loop approach focused around 3 pillars: raise awareness, support action, and foster engagement.

Towards raising awareness, we employ three directions: a) an online challenge, comprising quests and online games executed at individual and class level, b) a social networking game, prompting users to seek information, and, c) participatory sensing, which also provides knowledge hints.

Furthermore, in order to support action, a recommendation engine proposes steps towards better efficiency based on the data gathered from the school buildings. Such recommendations are addressed at the school building managers, as well as teachers and students. A dedicated user interface for the building manager and the teacher/students groups is utilized to fit their needs.

Complementing the above, engagement with GAIA is strengthened through:

- *Competition*, which is implemented in the online challenge: the communities of different educational facilities compete among each other and their scores are based on knowledge but also facility points earned upon reduction of energy consumption.
- *Social networking*, where the communities compete to respond to the challenge announced (i.e. find the next energy efficiency relevant event in EU).
- *Continuous monitoring* of energy consumption.

Including the users of a building in the loop of monitoring their daily energy consumption, creates the first step towards raising awareness. In an educational environment, it connects the educational activities carried out, with the students' activities at their home environment, as well as engaging their parents and relatives. The GAIA applications are designed offering new educational activities targeting preliminary/secondary/high schools and universities addressing the pedagogical needs of existing curricula in aspects related to the environmental impact of energy usage and environment-friendly/energy-efficient behaviors. Game-based methods are utilized to offer rich opportunities for student learning through engagement with the design and operation of the school building, providing a "synergistic relationship" [12] between enhanced student learning and an energy conservation focus.

While the previously outlined applications form a tool set that enable the design and implementation of a large number of energy efficiency campaigns that can suit the diverse targeted communities, it presents important challenges: it requires the installation of sensing devices for the collection of measurements regarding the energy consumption as well as the conditions (both physical and activity related), it requires the implementation of the applications in a manner that allows for customization per targeted community, it requires the use of diverse technologies towards developing this representative set that can reach all the communities members through a wide set of devices (desktop/laptops, mobile devices).

## IV. THE APPLICATION SET

To implement the proposed application set we have adopted the architecture shown in Figure 2.

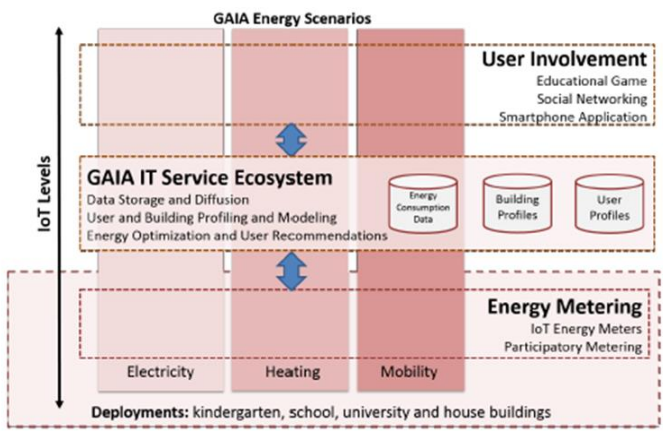

*Figure 2 Overview of the GAIA application set*

The GAIA cloud infrastructure is used by all applications since this infrastructure is used to store and retrieve facility and user-relevant data. The facility–relevant data include both static data like the building surface, types of energy used, etc. as well as real-time data collected from the sensing devices installed in the educational facilities. Optimization modules as well as other functional blocks are deployed in the GAIA cloud and offer their results (e.g., analytics, energy efficiency recommendations) as a service.

We now describe the applications emphasizing on the challenges they address.

### A. Educational serious game

The Educational Serious Game is an online challenge game which raises students' and teachers' energy awareness within their own facility and is accessible through web browsers. The challenge utilizes gamification mechanics to:
- motivate participants to engage in energy saving topics,
- work on online "quests",
- participate in real-life activities,
- experience their impact on the facilities' energy consumption over the course of the challenge,
- compete and compare against other classes and schools in other countries,
- share their experiences with their peer group.

Real-time data from sensors in the buildings will be used as an integral part of the challenge, in order to visualize a real life impact of the participants' behavior and build collaborative (within a facility) and competitive (between facilities) gamification elements upon the real-life impact.

The online challenge is modular enough in order to support the diverse communities (from elementary schools to secondary schools universities), diverse knowledge items (energy efficiency for different energy sources, controlling of the room climate resource sharing, etc.), diverse languages and cultures, diverse activities completed either on individual basis or an class/group basis, either digital (e.g., filling in multiple choice questions), or physical (e.g., monitoring the power consumption meters).

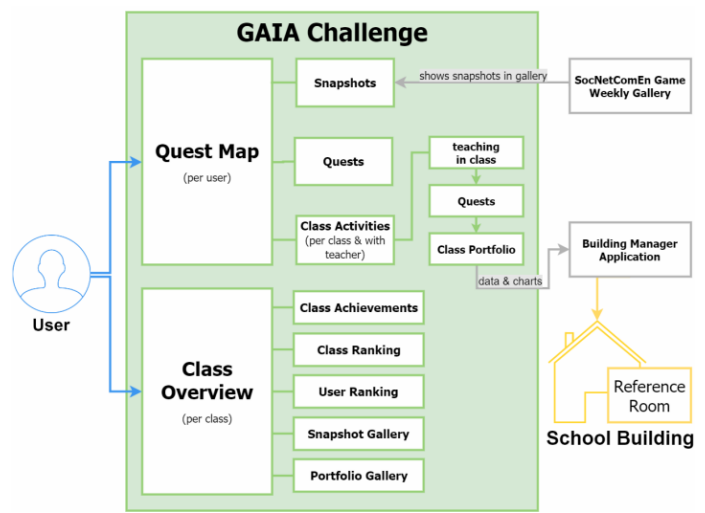

*Figure 3 Overview of the GAIA Challenge*

To meet these requirements local teachers are involved in Class Activities to work together with their students on hands-on observation and optimization tasks in class rooms. The online challenge offers the following core features:

*Quest Map:* The Quest Map is the main view for the user. It is an interactive visual representation of multiple entities of the online challenge. It symbolizes the user's journey from the start (top) to the finish line (bottom) of the challenge. Each Quest Node is related to a specific topic. Along the way there are also Quest Sequences and Class Activities, with multiple quests for the user to play. These quests are grouped into five subject areas related to energy consumption reduction. There are also bonus areas with quests available for classes that participate in Class Activities. The user has the option to submit a snapshot – a mixture of media content (photo, video) and a description (text) for each subject area. Snapshots are a way for the user to summarize and document experiences and findings.

*Class Activities:* Class Activities are crossovers of learning in class, engaging in the online challenge and on-site engagement in the facility, requiring a teacher and on-site engagement. The teacher can start a Class Activity for her class at any time during the online challenge. A Class Activity is divided into three parts: (a) learning the theory, (b) consolidating the knowledge and (c) applying it. Teachers can decide on the topic, the actual activities and the physical space (measuring the effect of hands-on optimization actions might require specific sensors) of these Class Activities. The results of the Class Activities are summarized by the class in the form of an online portfolio. Students and teachers are able to use the GAIA Building Manager Application in order to support their portfolios with data readings and charts. The submitted portfolios will be accessible in the online challenge and can be viewed by users and visitors. Users are also able to vote on their favorite portfolios.

*Community Dashboard:* users can compare the performance and score between classes and schools. The user can select a class from the list and inspect its achievements, its user ranking and also a list of recent snapshots by the class' students is shown. Visitors who do not participate in the challenge have to select a class from the list for further investigation. The class score is the sum of points of the class' students. Each student is rewarded with points for their performance on completed quests. The class badges are rewarded for special achievements such as reaching a high score in the quests.

### B. The Community Engagement and Social Networking Game

The aim of the Community Engagement and Social Networking Game is to provide an additional means to interact with the community and promote return/continued visits. For this reason, while the game's principle is the same across countries, more than one social networks are adopted, since different communities have different preferences, based on age group and country of implementation. Different groups per school are also created so that the different communities can have their own social networking communication channel to share information. The game carries user-generated content along with the gain they achieve, or possible suggestions towards awareness.

To accomplish this goal, the Community Engagement and Social Networking Game sets a weekly goal for the participating school communities towards fostering engagement and awareness. The game aims at raising awareness, sharing information, expressions of creativity and competition. It is a type of web scavenger hunt, using #hashtags for tracking purposes, as well as a place where the student-created content from the educational serious game can be shared. The players are asked within the App to do one of the following on a weekly basis: look for specific information or things, find an answer to a question, complete a certain task, create a unique image/GIF/video, or analyze and share something.

Once any of the tasks above have been completed, the player has the choice of either Tweeting or posting their answer on Facebook, along with that week's specific #hashtag. The #hashtags are tied to the player's score. These scores are a way for the players to keep track of how many scavenger hunts they have successfully completed. The information the players are asked to look for is completely relevant to the energy efficiency challenge put in.

### C. Application for Building Managers

The application for building managers is a responsive web-application (also accessible through mobile devices) offering direct visualization depicting in various forms and aggregation and analysis levels energy consumption information while also utilizing participatory sensing practices.

The building manager is able to a) insert building specific information, allowing for buildings comparisons and anomalies detection, b) insert data regarding building status (e.g. sensor readings or occupancy information) realizing participatory sensing, c) inspect building status and monitor building performance (building analytics), d) receive valuable suggestions for energy savings, e) communicate through social networks with other building managers and/or experts to get advice on energy savings. With respect to building inspection and monitoring, the building manager is able to inspect real-time energy usage where respective (energy and temperature) meters are available in various timescales (daily, weekly, monthly, yearly) as well as results from comparison with similar buildings or with the same building in other time spans (e.g. previous years) along with possible comments.

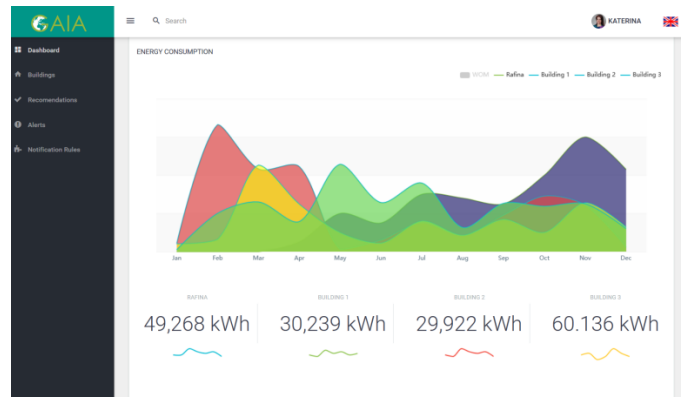

*Figure 4 A screenshot from the building manager application*

Participatory sensing is also supported for all actors of the school community. The building manager specifically is able to insert readings from sensors including (a) power meters not connected to the cloud, (b) fuel/heating system consumption, (c) luminosity, (d) indoor and outdoor temperature, (e) comfort level for luminosity and indoor temperature. The dashboard allows uploads of energy consumption in typical formats (manual entered monthly readings, file upload of hourly or 15-minute values) combined with meta-data on the building such as type, age, size, etc. It is worth noting that teachers and students can enter sensor readings or environmental and situational information in the platform. However, only building managers are capable of inserting the building data and configuring the facility setting.

The recommendations that the building manager will receive aim at improving energy efficiency of all energy types. Even within each category of building, the style of the recommendations may largely vary and so do the actions expected from the building manager and the building. The recommendations can be behavior-based, alerts and technical interventions, or building renewal actions.

*D. Recommendation Engine*

The aim of the Recommendation Engine is to generate appropriate recommendations for energy savings, according to the occurrence of some specific conditions in the building/area of interest (e.g. energy consumption peaks, absence/presence of users, temperature values, etc.), through a rule-based approach. Although the main aim of the module is to achieve energy savings through behavior-based changes, the module can support also the notification of additional types of advices (e.g., technical intervention, maintenance scheduling, etc.). Table I shows some basic examples of possible suggestions and related triggering conditions. More significant conditions could be modeled and detected by applying appropriate data mining techniques.

TABLE I. EXAMPLE OF RULE-BASED RECOMMENDATIONS

| Recommendation | Description | Condition |
|---|---|---|
| Turn-off the light | Turn-off the light when leaving | Room X is empty AND Light is ON |
| Standby | Do not keep electronic equipment (e.g., TVs, PCs) on standby when not in use | Lab X is empty AND Power consumption > threshold |

The rule engine has been developed in order to easily support the management of rules for different buildings and areas inside a building, as well as enabling the creation and modification of rules at runtime. It adopts a resource-based and RESTful approach for modelling and handling measurement values, as well as the rules to be applied. It is a Java-based application and uses OrientDB [16], which is a hybrid Document-Graph database, for persisting rules. Figure 5 shows an example of a tree of rules that can be specified for a school. The GUI offered by OrientDB can be used to easily navigate and modify the proposed resource-based rule model.

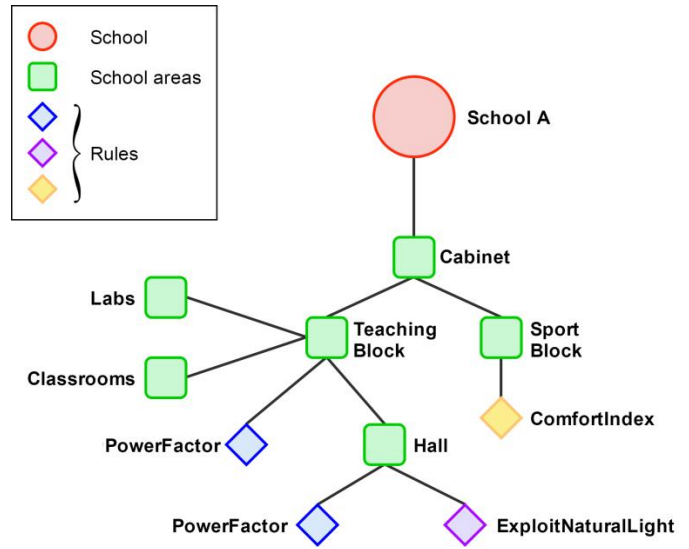

*Figure 5 GUI Resource-based rule model*

Once the occurrence of a condition specified in a rule is detected, the engine sends a notification message to interested applications (i.e., the Application for Building Managers) through a WebSocket channel [17]. Figure 6 shows an example of a notification message. The content of a notification message, which can be customized through the GUI mentioned above, includes the suggestion, i.e. a textual recommendation briefly describing what happened and how you may solve it, and a brief description of the event that triggered the rule.

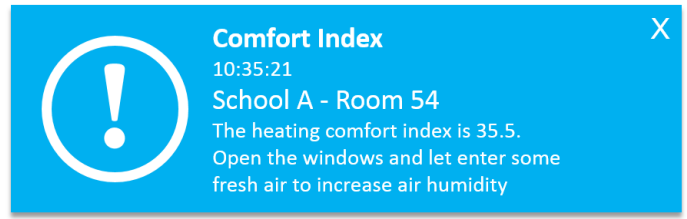

*Figure 6 Example of a Notification message*

## V. DISCUSSION

The value of GAIA's approach lies in: a) the variety of applications, that address all actors of the school community (already discussed), b) the flexibility of the applications (customizable to suit the specific facilities and communities intricacies), and, c) the openness in integrating additional applications in the future.

Shedding light at the adaptability of the applications, the online challenge (apart from supporting multiple languages) is open to integrate user-defined quests and challenges which renders it a flexible tool for any educational (or not) facility communities. Additionally, the building management application is flexible and novel: while the energy management and building management system market is wide and offers a great gamut of solutions, these remain expensive and require significant investment. Moreover, once installed, their upgrade after several years may incur vendor lock in. Instead, our approach follows a more open approach. Our platform remains open to sensing and actuating devices from a large set of vendors (it supports multiple different interfaces) and supports building managers' decision making. Furthermore, the capability to upload measurements in the form of individual values, or files, enables comparisons with historic building data, as well as with data from other similar buildings.

For the validation of the GAIA IT services and applications, a set of trials is currently being planned in a number of educational facilities across Europe (Greece, Italy and Sweden), representing different climate zones, building age categories and socio-economic criteria. In terms of number of buildings and people directly involved in this process, over 15 buildings and several thousands of students and educators will participate in the activities envisioned.

This series of trials in schools and universities will evaluate the different kinds of feedback developed, assessing which is the better way for each user category (and age) to perceive energy amounts or its energy spending profile. An experimental-driven research approach will be adopted, one that requires the development of a "living lab" to involve users at an early stage in the design process for "sensing, prototyping, validating and refining complex solutions in multiple and evolving real life contexts" [13]. All the pilot buildings are continuously collecting energy consumption and environmental data, gradually gathering a detailed profile in terms of energy consumption and other characteristics, which will also help to provide in the future a group of datasets for the research community.

## VI. CONCLUSIONS

Energy efficiency at the educational domain is quite significant due to the fact that it reserves a large portion of the budget for school buildings, while in most cases such buildings were built over 20-30 years ago, leaving considerable space for a large number of optimizations, [14, 15]. In many case studies available, interventions regarding energy use in school building operations led to sustainable best-case savings below 15%. The GAIA Application Set aims to accelerate the reduction of the systemic energy consumption and production of emissions resulting from educational buildings. As described in this work, GAIA aims to engage the students by utilizing bidirectional communication with the educational community, gaining valuable feedback that leads to more efficient intervention for changing students' habits regarding inefficient energy use. We believe that a participatory, ethical and sustainable engagement of students is a meaningful way to enable greater understanding of these concepts and lead to future well-informed, responsible and active citizens. With respect to our future work, GAIA will put to the test the application set described here through trials conducted in a number of EU school buildings.

ACKNOWLEDGMENT

The GAIA Application Set is part of the Green Awareness in Action (GAIA) project funded by the European Commission under the H2020-EE-2015-2-RIA call and contract number 696029.